\documentclass[aps,pre,groupedaddress]{revtex4-1}

\usepackage{graphicx}

\usepackage[english,francais]{babel}
\usepackage{amssymb,amsmath}
\usepackage{subcaption}
\newtheorem{e-proposition}[theorem]{Proposition}

\newtheorem{e-definition}[theorem]{Definition\rm}

\renewcommand{\theequation}{\arabic{equation}}
\def\og{\leavevmode\raise.3ex\hbox{$\scriptscriptstyle\langle\!\langle$~}}
\def\fg{\leavevmode\raise.3ex\hbox{~$\!\scriptscriptstyle\,\rangle\!\rangle$}}

\begin{document}

\title{Pattern Universes}

\title{Pattern Universes}
\author{Alan C. Newell}
\email{anewell@math.arizona.edu}
\thanks{Pierre Coullet has been both a friend and inspirational colleague for over thirty five years. Besides pioneering many of the early advances in dynamical systems, he has carried the ideas of universal behaviors from the ode to the pde world and in particular to pattern forming systems. With tremendous imagination and with his belief that physical ideas when appropriately presented can be understood by children of all ages, he has played a central part in making physics accessible and exciting. It is an honor and a privilege to be part of this meeting and this volume.}
\affiliation{Department of Mathematics, University of Arizona, Tucson, AZ 85721}

\author{Shankar C. Venkataramani}
\email{shankar@math.arizona.edu}
\affiliation{Department of Mathematics, University of Arizona, Tucson, AZ 85721}

\date{\today}


\selectlanguage{english}

\begin{abstract}
In this essay we explore analogies between macroscopic patterns, which result from a sequence of phase transitions/instabilities starting from a homogeneous state, and similar phenomena in cosmology, where a sequence of phase transitions in the early universe is believed to 
have separated the fundamental forces from each other, and  also shaped the structure and distribution of matter in the universe. We discuss three distinct aspects of this analogy: (i)  Defects and topological charges in macroscopic patterns are analogous to spins and charges of quarks and leptons; (ii) Generic (3+1) stripe patterns carry an (energy) density that accounts for phenomena that are currently attributed to dark matter; (iii) Space-time patterns of interacting nonlinear waves display behaviors reminiscent of quantum phenomena including inflation, entanglement and dark energy. 

\vskip 0.5\baselineskip

\selectlanguage{francais}
\noindent{\bf R\'esum\'e}
\vskip 0.5\baselineskip
\noindent
{\bf Univers de structures.}
Dans cet article, nous explorons plusieurs analogies entre la formation de structures p\'eriodiques macroscopiques, qui r\'esultent de la succession de transitions de phase 
ou d'instabilit\`es, et certains ph\'enom\`enes similaires en cosmologie, o\`u une suite de transitions de phase dans l'univers primordial aurait donn\'e lieu \`a la s\'eparation des forces fondamentales et \`a la formation des structures. Nous consid\'erons trois analogies diff\'erentes: (i) Les d\'efauts et charges topologiques dans les structures macroscopiques sont analogues aux spins et charges des quarks et des leptons; (ii) Les structures p\'eriodiques g\'en\'eriques (en dimensions 3 + 1) ont une densit\'e d'\'energie qui donne lieu \`a certains ph\'enom\`enes attribués \`a la pr\'esence d'mati\'ere noire; (iii) Les structures spatio-temporelles r\'esultant de l'interaction d'ondes non lin\'eaires ont des comportements qui rappellent certains ph\'enom\`enes quantiques, tels que l'inflation cosmique, l'enchev\^etrement quantique, et l'\'energie noire. 


\end{abstract}

%
%
%


%
%
\maketitle

\selectlanguage{english}

Our aim in this essay is to introduce a new paradigm for thinking about the evolution of the universe. Building on previous work on patterns and pattern formation, we suggest  a new theoretical construct modeling aspects of the universe on a wide range of scales, from the very small (Planck scales) to the very large (galactic scales).  Our ``toy models" display parallels to many of the challenges to and mysteries of current thinking in cosmology. We certainly do not claim that what we suggest resolves all the challenges to current theories. Rather, we present an alternative viewpoint with which to approach these challenges. While it may or may not have anything to do with the real universe, it nevertheless has several intriguing parallels. 

We label our universe as a pattern universe as it basically sees the evolution of space-time, matter-energy, particles with charge and spin as natural and universal outcomes when, under changes in various stress parameters, systems undergo phase transitions from one pattern forming state to another. At each phase transition, a new pattern or macroscopic state is formed. For example, under a change of temperature difference between parallel, horizontal plates containing a layer of fluid, the purely conductive state (basically thermal noise with a temperature gradient) gives way to a state consisting of buoyancy driven cells which carry additional heat from one plate to another via convecting cells in an almost periodic structure with a preferred wavelength. The new macroscopic state is described by variables called {\em order parameters} which, in the case of the cited example, would be the slowly varying amplitude and phase of the underlying almost periodic pattern of convective rolls. The physics of the pre-transition stage are governed by one set of variables whereas in the post-transition they are governed by the dynamics of the new order parameters superposed on the preferred almost periodic structure. The two states may reflect very different scales and exhibit very different behaviors. The pattern universe we envision will undergo many such transitions, from a primordial state of fluctuations, a sea of waves, to a more ordered state governed by equations one might associate with Einstein's equation of classical relativity whose properties such as the speed of information travel of macroscopic information can be very different from speeds at which the fluctuations of the pre-transition state may travel. Although the initial transitions will involve the Planck length and time scales, later transitions will reflect all the other instability driven transitions occurring at longer and longer length and time scales. The largest of these, which gives rise to the pattern which contains the energy which provides the pattern parallel of dark matter, is the gravity induced clustering instability generally attributed to Jeans. 

What is important to note is that the relevant variables at each stage, the order parameters, which describe the behavior of each new emerging state, satisfy {\em universal} and {\em canonical} equations.  These equations  do not depend on all the microscopic details of the original unstable system but rather only on its underlying symmetries. We can thus, oftentimes, write down the corresponding order-parameter equations without a detailed analysis of the underlying microscopic models. As a consequence, the nature of new laws at each new level of scale depends very little on details of the laws at earlier levels. This {\em principle of universality} is indeed a cornerstone for studying condensed-matter/pattern forming systems, and our goal in this essay is to explore possible applications of this principle to the universe. 

Beyond a phase transition resulting in a new preferred state,  the order parameter equations describe  solutions corresponding to a mosaic of patches of periodic or quasiperiodic patterns reflecting the new preferred state separated  by lower-dimensional defects. In our pattern universe model, defects carry energy, which translates into hidden mass, as well as  topological indices which provide analogues of spin and charge. The order parameters capture the envelope or averaged properties of the new state, both its smooth regions and its defects, and, as we shall see, exhibit many features and behaviors which parallel the manifestations of dark matter, energy and particle structures.

Faced by substantial observational evidence that the universe, or at least the portion of it we can experience, is flat (BOOMERanG \cite{boomerang} and WMAP \cite{wmap}) and weakly accelerating in its expansion \cite{Peebles_The_2003} and that the rotation of stars in galaxies and galaxies about clusters is not consistent with the detectable mass content (Rubin et al  \cite{Rubin_Rotation_1970}), current cosmological thinking has had no alternative but to posit the existence of {\em dark energy} and {\em dark matter}. Let us be clear. Neither designation explains anything. For the moment, they are simply descriptive place holders until the reasons for a new kind of pressure (dark energy) can be elucidated and a new kind of matter (dark matter), that only interacts with the observable universe via its gravitation, can be identified. The amounts of each required to reconcile with observations stagger the mind. Dark energy contains about 70\% of the energy content of the universe. The remaining 30\% consists of (roughly) 25\% dark matter and 5\% all other detectable matter. So 95\% of all the stuff from which the universe is composed has origins we simply do not yet understand while the stuff we can understand, neutral and ionized matter, bricks, walls, you, me, the earth, the stars, is only 5\% of what is out there. Therefore, new ideas as to the possible origins of mysterious dark energy and matter should be welcome.

What we suggest also has vast holes of ignorance but it has (in our view) one key advantage -- it represents both dark matter and dark energy as manifestations of behaviors that we already know from studies of macroscopic pattern formation. It also has the merit that 
the building blocks of observable matter, namely quarks and leptons with fractional charge and spin, have pattern analogues.

As one might expect, this paradigm has many things in common with current thinking. The pattern universe starts out as a sea of fluctuations with very high values of some stress parameter such as temperature. Nonlinear interactions in this sea spread its energy throughout a spectrum of scales and the relative speeds of modes of different sizes can be arbitrarily fast. The speed of light may limit travel through space but the space-time manifold itself can expand in any way it wants. One might liken the rapid expansion as to that experienced by a turbulent ball or by wave turbulent eddies in classical systems. In this sea, the dominant scales are of the order of the Planck length. As the stress parameter, temperature, lowers, a series of phase transitions takes place, each one highlighting a set of new structures which emerge as energy minimizers in the new state. They bring about but need not end with analogues of gravity and other forces but may involve patterns on patterns with scales up to and including galactic scales. The order parameters of each new state are the envelopes of the most excited modes and modes which are called soft or Goldstone modes, which reflect some symmetry of the original system and which are neutral at the phase transition. 

In pattern theories, we know that the linear parts of these envelope equations reflect the dispersion relation which obtains to describe the onset of new modes emerging as instabilities of the former state. They have universal character. For example, in a convection pattern dominated by stripes of a particular orientation, the order parameter might be the envelope of the most unstable mode and its nearby wavenumbers, and the soft mode (relevant in convecting fluids with moderate to small Prandtl numbers) might reflect the pressure (which can change by a constant without changing the equations) or vertical vorticity (in the absence of viscosity). The resulting partial differential equations are basically Taylor series expansions to the lowest nontrivial order in the (complex valued) envelope function $W$, its spatial and temporal derivatives, and, at least near onset, products of the amplitude with the difference between the stress parameter and its critical value. They may also contain forcing terms, originally called geometric imperfections by Koiter, which nowadays are more commonly referred to as imperfect or continuous bifurcations. These pre-existing biases can smooth a phase transition by prematurely forcing the mode that is about to become unstable.

 It is very important to stress again that the order parameter equations are canonical and universal, expressing symmetries in the state about to be destabilized. Vastly different microscopic systems sharing symmetries will have very similar equations for their order parameters. Only in the choices of coefficients are the details of the microscopic system remembered. 

As the difference between the stress parameter and its critical value grows, the amplitude of the envelope usually becomes slaved to gradients of its phase gradient, so that the order parameter equations become a description of how the phase of the pattern evolves, coupled to any soft mode effects. The slaving of the amplitude has extremely important and relevant consequences. When the order parameter is a complex variable, $W=A\exp(i \psi)$, as it is near the phase transition when both the amplitude $A$ and the phase $\psi$ are active order parameters, the phase gradient can be determined uniquely from a knowledge of $W$ or its underlying microscopic variable $w$. It is therefore a vector field. However, far from equilibrium, when the relation between the original field and the order parameter is $w=f(\psi;A)$, with $f$ a $2\pi$ periodic function in $\psi$,  where the amplitudes $A$ are slaved to phase gradients, then the phase gradient can only be determined by the original field $w$ up to sign. Therefore, whereas in the former case, the wave-vector or phase gradient order parameter is single valued, in the latter case it is double valued. Namely, far from equilibrium the phase gradient order parameter is a director field, such as that commonly and necessarily used in the description of liquid crystals. This fact greatly changes the topological character of the allowed defects. For example, for director fields the point defects in two dimensions are disclinations ($X$ and $V$) with topological indices (see Appendix~\ref{apndx:patterns}, Fig.~\ref{point-defects}) of $\pm \frac{1}{2}$ with fractional spin character. For vector fields, they consist of composites of disclinations, such as targets ($XX$), vortices ($XX$ in a different configuration), saddles ($VV$), handles ($VX$) and dislocations ($VVXX$), all of which have integer spins. In the picture we discuss, spin $\frac{1}{2}$ particles are associated with director fields, whereas integer spin defects such as point masses will occur as composites and defects in vector fields. For composites, the circulation, namely $\int \vec{k} \cdot d\vec{x}, \vec{k} = \nabla \psi$ is well defined in addition to the topological indices, called twists, which give rise to half integer spins.

Current cosmology theories tend to think of and talk about the evolution of the universe in terms of preferred energy states assuming thereby that the dynamics is governed by a gradient or Hamiltonian flow with an associated free energy or Lagrangian/Hamiltonian. While, for the most part, we also adopt that viewpoint, we are cognizant that very few flows in nature are gradient flows. Indeed, for most parameter ranges in convecting fluids, the observed states do not minimize an energy functional at all with the consequences that, while relatively ordered, they display patterns with continuous and chaotic time dependence and patch/defect structures that do not necessarily coarsen. A good example is the spiral defect chaos consisting of contiguous spirals in convecting flows at moderate Prandl numbers. Therefore, while we will find it convenient to think of the patterns which evolve in pattern universes as arising from minimum energy states, we will also keep it firmly in mind that the governing equations for the phase dynamics may or may not arise from energetic considerations.

In Appendix~\ref{apndx:patterns}, we introduce in two and three dimensional systems the energy functional governing the phase of the pattern. Our goal here is to exhibit pattern analogues of quarks and leptons. The energy functional consists of two terms, one arising from the first (by analogy with elasticity we call it the strain energy) fundamental two form of the phase surface $\psi$, namely the metric form, and the second, after times long compared with the horizontal diffusion time, arising from the second fundamental or curvature two form. In each case, the combinations of the metric and curvature tensor which appear are invariant under Euclidean coordinate transformations. As the pattern evolves on the horizontal diffusion time-scale, the time it takes for macroscopic information to travel across the pattern, the first term which is initially dominant relaxes to a state where almost everywhere the pattern has a preferred structure (for the purposes of this discussion, stripe-like, although it can also be hexagonal, or even quasi-periodic) and a preferred wavelength $2 \pi/k_0$. Deviations from the preferred wavenumber $k_0$ occur along defects and in these regions the curvature part of the energy, which by this stage has canonical form only depending on the curvature tensor, balances the strain part. 
The curvature energy consists of terms in its integrand proportional to the mean curvature and all the sectional Gaussian curvatures. The latter however integrate out to boundary terms and are then determined by whatever topology is imposed by the far field. For example, in the region of the concave disclination, shown in Fig.~\ref{concave} of Appendix~\ref{apndx:patterns},  the Gaussian curvature $\psi_{xx} \psi_{yy}-\psi_{xy}^2 $ integral is given by $\frac{1}{2}k_0^2$ times the twist of the director field on a contour surrounding the point defect\footnote{Caveat: we choose to call $\psi_{xx} \psi_{yy}-\psi_{xy}^2 $ the Gaussian curvature when in fact we should define the latter as this quantity divided by $(1+\psi_x^2+\psi_y^2)^2$ but it can easily be shown that this too can be written in divergence form and again integrates out to a multiple of the twist.}. The topologies imposed carry information of pattern charges, spins and masses.

In a very real sense, then, everything in pattern universes is geometry, and in particular the sectional Gaussian curvatures of the phase surface generated at each phase transition. In situations where flows are gradient, energy minimization is achieved by the juxtaposition of two equations, the variation of the energy with respect to the phase and the variation with respect to the Lagrange multiplier describing the constraint which equates the sectional Gaussian curvatures to the twists due to boundary or far distance conditions. In many cases where the Gaussian curvatures condense onto points
(we will describe the general case later), the variation with respect to the phase gives an equation of fourth order which can be  ``solved'' by adding an extra term to the self-dual solution. The extra term then is determined by an equation whose forcing is provided by the nonzero Gaussian curvatures. When they are condensed onto points, this term acts as a Dirac delta function which modifies the self-dual solution along defect surfaces, curves and points. In regions where the Gaussian curvature is zero, the self-dual solution, in which the integrands involving mean curvature and strain energy are equal, obtains. For example in two dimensions, namely when the phase surface is a two dimensional surface embedded in three space, if the shape of the phase surface is planar or cone-like, then the Gaussian curvature is zero and the self-dual balance obtains. In the two dimensional point defects shown in Fig.~\ref{point-defects}, the Gaussian curvature is zero except at the point defect itself. In the three dimensional loop defects shown in Fig.~\ref{loop-defects}, the sectional Gaussian curvatures are confined to the backbones. The topological indices, the analogues of charge and spin, are measures of the total amounts of sectional Gaussian curvatures which the loops circumscribe. The loop disclinations which we call pattern quarks and leptons are analogous to the more familiar vector field loops with vortex cores seen in fluids (smoke rings) and in organic tissue where they are often called scroll waves.

In appendix~\ref{apndx:stress-energy}, we describe the pattern analogue of dark matter. When natural patterns first form, it is rare that a simple, energy minimizing planform covers the whole domain. While at the phase transition, certain symmetries are broken, for example the symmetry of translation in a field of convecting rolls, other symmetries remain. For example, in a horizontal layer of fluid heated from below with both translational and rotational symmetry, while the planform of parallel rolls is chosen with a preferred wavelength thereby breaking translational symmetry, the orientation of a patch of rolls depends on local biases such as boundary conditions or initial fluctuations in the system. As a result, natural patterns arise as patches of the preferred planform with an orientation chosen by local biases and these patches meet and meld along line and point defects. The patches will eventually coarsen but this process takes a very long time (which can be quantified). Moreover, eventual coarsening assumes that the system is governed by gradient dynamics but most systems we meet in nature are not gradient systems at all. In such cases, there is no requirement that defects eventually coalesce and disappear. Therefore, we can expect that the pattern universe will exhibit the same features, namely that it will contain lots of defects, either very long or infinitely long lived as it evolves. When we talked about pattern quarks and leptons, we learned about the character of these defects, their invariant indices (charges and spins) and how these indices are related to geometry and in particular the condensation of Gaussian curvature of the phase surface onto these same defects. But the defects also contain energy as the system is not in a total minimal energy state. What we demonstrate in Appendix~\ref{apndx:stress-energy} is that the energy contained in a class of integer spin defects, specifically targets, gives rise to an effective mass and corresponding force. This force will act in galaxies and clusters of galaxies as an additional component in the net gravitational force experienced by stars in a galaxy and galaxies in a cluster.

We will show that the manifestation of this force has an equivalent effect to that of the dark matter posited by traditional cosmology. Therefore, our contention is that, in a pattern universe, dark matter is simply a manifestation of the energy contained in pattern defects.  Furthermore, it gives a result for the dependence of the rotation velocities of stars on radius which matches closely (as shown in Fig.~\ref{fig:rotation}) what has been observed by Vera Rubin and colleagues. Namely, in the far field, the rotational velocities of stars in galaxies and galaxies around clusters tends to a constant. 

In Appendix~\ref{apndx:quantum-patterns}, we discuss a pattern analogue for dark energy, that mysterious pressure force which increases as the mass density decreases. In our pattern universe, we think of the original microscopic state as involving fluctuations with time and space scales of Planck time and length. The first phase transition picks out what we call the gravity force and just as in the cases of pattern quarks and leptons and dark matter, its envelope phase equation reflects the sectional Gaussian curvature of the space-time manifold and has a form similar to Einstein's equation. In the weak mass limit where the metric tensor $g_{\alpha \beta}$  is a small perturbation $g_{\alpha \beta} = \eta_{\alpha \beta} + h_{\alpha \beta}$ of the Minkowski metric $\eta_{\alpha \beta}$, the order parameter equation is a weakly nonlinear wave equation for $h_{\alpha \beta}$. The mass density in our picture arises from both the bulk and the defects. The weakly nonlinear terms arise from two sources, a self-interaction and the interaction between the wave and a Goldstone or soft mode. It is the soft mode which resembles the cosmological constant. It is not quite constant as it reflects some global average of the order parameter intensity. Our picture is that behaviors on all scales can be accommodated within a single theory although the behaviors at different scales will depend on the behaviors of the phase surfaces generated by successive transitions. The short scale stuff, the Planck scale ``quantum'' part, is represented by the underlying periodic pattern. Its envelope, depending on much longer space and time scales, carries information on the classical level. In later transitions leading to analogues of the strong, weak and electromagnetic forces, the envelope equations should not only capture the physics on classical scales but should also encode quantum behaviors on atomic and nuclear scales. 

In pattern forming systems, the appearance of a wave equation is very natural for the post instability stage of a transition in which a subset of fluctuating modes become amplified as almost stationary states. The second time derivatives naturally arise because square of the growth rate $\sigma$ goes from negative to positive at the phase transition and the mode with wave-vector $k$ which maximizes the growth rate $\sigma^2(k)$ is the preferred one\footnote{If the growth rate has a simple zero at the phase transition, the usual Ginzburg-Landau or complex Ginzburg-Landau equation obtains}. In other words, the transition is from waves to stationary growth. Again, the order parameter equation for this class of transitions has universal character independent of details of the microscopic state and only dependent on its symmetries. Therefore, the precise details of how physics works at the quantum levels does not greatly affect behaviors on classical time and length scales. However, the microscopic details are remembered by the coefficients in the universal envelope equations. They appear as functionals of parameters which appear in the microscopic state and express what might be called fine structure relations.

In Appendix~\ref{apndx:quantum-patterns}, we introduce a very simple model to illustrate how, close to onset, the order parameter equations for the envelope of the most unstable mode and a soft or Goldstone mode are wavelike with the soft mode parameter playing the role of a cosmological constant. Working from this model, we discuss several possible behaviors. First, we explain that in this picture, the cosmological constant is a field rather than a constant of fixed value. Second, we discuss how it can play a role of the pressure in an equivalent hydrodynamical description, a pressure which increases with decreasing density. Third, we note that its interaction with the wave field can be both passive or active in the sense that changes in the latter can produce large effects in the soft mode field. While the possibilities might be enticing, we stress that they are just that, possibilities, and vague ones at that. Their connections with any concrete behaviors of the real universe are highly speculative. Moreover, we have no suggestion as to why any transition in the universe's evolution should have any soft mode at all or to which symmetry it might possibly correspond. 

As we have repeatedly said in the early parts of this essay, our aim is not to claim a new theory but simply to introduce a new perspective through which lens the evolving universe might be viewed.

\appendix

\setcounter{equation}{0}
\renewcommand\theequation{\Alph{section}.\arabic{equation}}

\section{Pattern quarks and leptons} \label{apndx:patterns}

The results we discuss here have already been extensively reported in several earlier publications \cite{newell2012pattern,newell2014quarks,NV18} but some main details are worth including here because we want to consider them as part of a wider story. In the previous work, we had introduced three dimensional loop disclinations as extensions of two dimensional concave and convex point disclinations on closed loop backbones. 

\begin{figure}[h!]
    \centering
    \begin{subfigure}[b]{0.25\textwidth}
        \includegraphics[width=\textwidth]{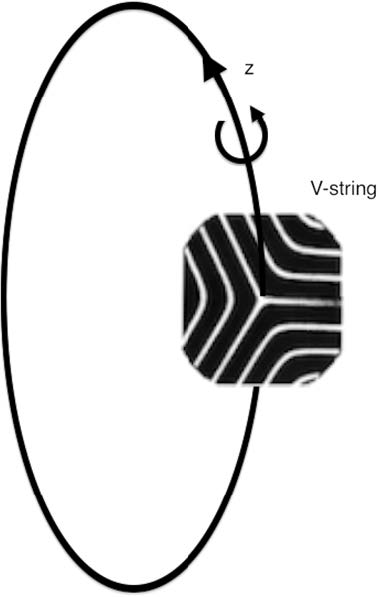}
        \caption{A V-string.}
    \end{subfigure}
    ~ 
    \begin{subfigure}[b]{0.25\textwidth}
        \includegraphics[width=\textwidth]{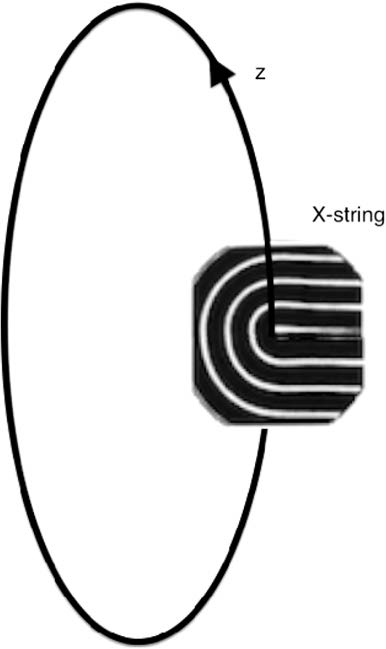}
        \caption{A X-string.}
    \end{subfigure}
    \caption{Loop defects. $z$ is the coordinate along the ``backbone" and the pattern is periodic in $z$.}
    \label{loop-defects}
\end{figure}

Associated with such objects were two topological indices which we termed spin and charge and since the indices turned out to be integer multiples of 1,1/2 and 1/3 we called them pattern quarks and leptons. 
We briefly summarize some background. We begin with a simple example of a two space dimension microscopic field evolving under the dynamics
\begin{equation}
\label{Swift-Hohenberg}
w_t  = R w - (\nabla^2+1)^2 w - w^3  = - \frac{\delta E}{\delta w}
\end{equation}
for the real valued scalar field $w(x,y,t)$, where $R$ is the stress parameter. More complicated equations with the same symmetries will yield similar averaged properties. We seek slowly modulated solutions
\begin{equation}
w=f(\psi; \{A\},R) + \text{corrections} + \ldots ,\quad f  \text{ is  }2 \pi \text{ periodic in } \psi                      \end{equation}
where the amplitudes $\{A\}$ are slaved to the phase gradient $\vec{k}= \nabla \psi$, a director field. By averaging \eqref{Swift-Hohenberg} over the period $2 \pi$, or by finding the solvability condition for the higher order corrections in $w$, we obtain the phase diffusion equation,
\begin{equation}
\label{CNreduced}
\langle w^2_\psi \rangle \frac{\partial \psi}{\partial t} = - \frac{\delta}{\delta \psi} \left\{ \int \frac{1}{4} \langle w^4 \rangle \bigg|_{k^2}^1 + \epsilon^2 \langle w^2_\psi \rangle \left((\nabla_{\mathbf{x}}\cdot \mathbf{k})^2 - 2(\psi_{xx} \psi_{yy} - \psi_{xy}^2)\right) dx dy \right\}.
\end{equation}
In the early stages of the relaxation, in convection terms the horizontal diffusion time, the first or strain term in the average energy $F$ dominates but as it does the wavenumber $k$ almost everywhere approaches its preferred value $k_0$, in this context equal to 1. Only on defect lines and points will it significantly differ and in these regions it is balanced by the curvature energy.

\begin{figure}[bth!]
    \centering
    \begin{subfigure}[htbp]{0.25\textwidth}
        \includegraphics[width=\textwidth]{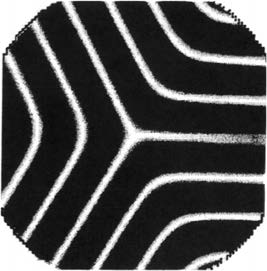}
        \caption{Concave disclination.}
        \label{concave}
    \end{subfigure}
\quad     
    \begin{subfigure}[htbp]{0.25\textwidth}
        \includegraphics[width=\textwidth]{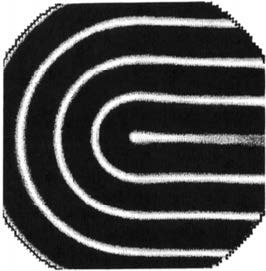}
        \caption{Convex disclination.}
        \label{convex}
    \end{subfigure}
       \caption{Point defects in 2D patterns}
        \label{point-defects}
\end{figure}

At first sight, it may seem superfluous to include the Jacobian (determinant of the Hessian) term in the integrand of the curvature energy as it has divergence form and can therefore be replaced by a boundary line integral 
\begin{equation}
J = \psi_{xx} \psi_{yy} - \psi_{xy}^2, \quad
2 \int_\Omega J(x,y) dxdy = \oint_C k^2 d \phi,
\label{eq:total_twist}
\end{equation}
where $\vec{k}=(k \cos \phi,k \sin \phi)$ and $C$ is a counterclockwise contour on the double cover of the plane. Moreover, the variation of $F$ with respect to the phase $\psi$ is zero. Nevertheless, it is very important because the relation~\eqref{eq:total_twist} manifests the influences of singularities in the phase field within the integration domain and represents as a consequence a constraint of the director field. For the two dimensional patterns shown in 
Fig.~\ref{point-defects}
of a concave and a convex disclination, the twists are $-2 \pi$  and $+2\pi$ respectively, which, when divided by the double cover angle $4\pi$, are $\mp \frac{1}{2}$ respectively. The angular separations of $2 \pi/3$ in the concave disclination (Fig.~\ref{point-defects}(a)) are a consequence of energy minimization. By similar arguments, in a three dimensional pattern with pattern singularities loop concave and convex disclinations as shown in Fig.~\ref{loop-defects}, there is a second class of twists or topological invariants representing the director field twist around contours twisting around the loop backbones. It is relatively easy to see these are integer multiples of $1/3$ and $1$.

Therefore, the constraints on the phase surface imposed by the natures of the pattern singularities involve an additional equation to the variation of $F$ with respect to $\psi$, which involves a relation between the mean curvature and the deviation of the pattern wavenumber from its preferred value. The two equations are compatible. The first is an integral constraint reflecting the nontrivial topology of the director field within the integration domain. The second describes the structure of the field in the neighborhood of defects.

There are two conclusions we want to stress. The first is that, without any a priori imposition of symmetries on the original microscopic field $w$ other than translational and rotational invariance, there nevertheless appear, entirely naturally without any imposition of $U(1), SU(2), SU(3)$ and $SU(5)$ symmetries, objects with topological invariants which are integer multiples of the fractions $\frac{1}{2}$ and $\frac{1}{3}$. 

The second is that we can interpret the integral constraint on the Jacobian $J$ (or Gaussian curvature) of the phase surface as being a constraint on the sectional Ricci curvatures of the manifold obtained by embedding out phase surface $\psi(x,y)$ or $\psi(x,y,z)$ in the three or four dimensional space $x,y,s$ or $x,y,z,s$ with the standard Euclidean metric $dx^2+\ldots+ds^2=(1+\psi_x^2)dx^2 + \ldots$. Namely, the integral constraint is the integration of the Ricci tensor or ``Einstein's equation'' over the domain. It might also be written $J=\rho(x,\ldots)$ where $\rho$ is some density. In our case here, of course, with the pattern structures shown in Figs.~\ref{loop-defects}~and~\ref{point-defects}, the density $\rho$ is a distribution. Indeed, this is entirely what we expect. As we have said, the relaxation of the pattern to its lowest energy state comes in two stages. The first occurs on the horizontal diffusion time scale (proportional to $1/\epsilon^2$ where $\epsilon$ is the ratio of pattern wavelength to the total pattern size containing many wavelengths) and in that stage the wavenumber almost everywhere takes on its preferred value. If that is true, then almost everywhere $f^2+g^2=k_0^2$ where $\nabla \psi=(f,g)$. It is easy to show that the Jacobian $J=f_xg_y - f_yg_x$ is also zero. Therefore, we expect $J$ to be distributed on defects, although not necessarily always on points as in Fig.~\ref{point-defects}. 

Indeed, if one takes a slightly more complicated geometry such as that of a semicircle domain with ``heated'' boundary conditions, namely the wave-vector $\vec{k}$ is perpendicular to the boundary, then a curve of wave-vector discontinuity forms in its interior. The usual line defect with a sudden but smooth wave-vector transition in a boundary layer is unstable because the constant phase contours intersect the curve defect at too sharp an angle and the final state is that of a dense set of dislocation like points along this contour. However, outside of this contour, the wavenumber $k$ attains its preferred value. For a discussion of the instability that leads to concave-convex disclination creation from a smooth vector field, see~\cite{NV18}. That source also describes how, near critical values of the stress parameter, such objects can spontaneously collide and restore the vector field nature of the wave-vector field.

Far from onset, therefore, we suggest that the minimization of the energy leads to a phase surface where the boundary layer structures satisfy $\delta E/\delta\psi=0$ and whose Jacobian $J$ is concentrated, namely it is nonzero only on a set of small measure. In this case, as we have shown in \cite{newell1996defects}, we can split the fourth order equation for the phase into a pair of second order equations by setting the mean curvature, the surviving integrand of the curvature energy, equal to the strain energy integrand plus a correction. The correction then obeys a very similar second order equation with a forcing term proportional to the Jacobian $J$, which we can take to be given based on the kinds of singularities we expect the pattern to have.

Before we leave this appendix, however, we list several some of the outstanding challenges connected with point and loop disclinations. For examples: Can we show that the interaction forces between such objects confirm their interpretations as charges? What is the nature of their composites? Do loop disclinations form knot like structures or do they, like smoke rings, tend to merge separate loops into a single one?  

\setcounter{equation}{0}

\section{A patterns inspired model for dark matter} \label{apndx:stress-energy}

\begin{figure}[htbp] 
   \centering
   \includegraphics[width=0.6 \textwidth]{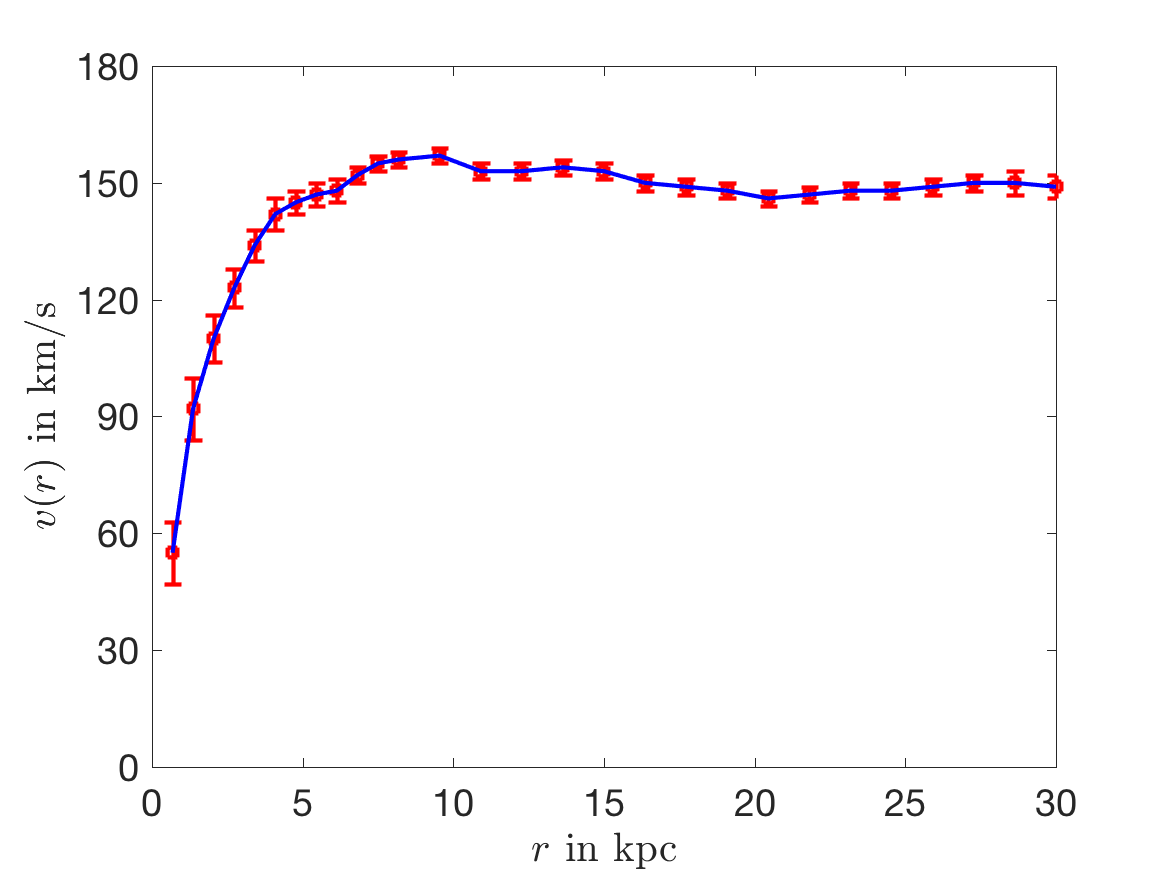} 
   \caption{The rotation curve for NGC 3198. $r$ is the distance from the galactic center and $v(r)$ is the rotation speed. The data is from van Albada {\em et al} \protect{\cite{van_Albada_1985}}.}
   \label{fig:rotation}
\end{figure}

We begin by offering the idea that a pattern universe has an extra energy which could explain the effects 
which have led to the suggestion from cosmologists \cite{Blumenthal_Formation_1984,Davis_Evolution_1985} that, in addition to visible matter, there exists a significant source of gravitation whose mass is many times that of visible matter \cite{Zwicky_Masses_1937,Rubin_Rotational_1980}. Important findings which led to a widespread acceptance of this conclusion are the studies of Rubin, Ford and Thonnard~\cite{Rubin_Rotation_1970,Rubin_Extended_1978} on the rotational speeds of stars in galaxies.  The simple model in which the force experienced by the star executing a circular orbit at radius $r$ about the galactic center due to some large center mass $M$ of the galaxy is balanced by the star's centrifugal force leads to a conclusion that the rotational velocity $v$ should decay with increasing distance $r$. This is contradicted by observations that consistently demonstrate that the galactic rotation curves flatten out and the orbital velocity of distant stars is roughly constant. This is illustrated, for example, in fig~\ref{fig:rotation}  generated from rotation velocity data for the galaxy NGC 3198 (van Albada {\em et al} \cite{van_Albada_1985}).

Turning the argument around by balancing $GM/r^2$ with $v^2/r$ with $v$ constant, gives a mass $M$ of $v^2 r/G$ which is more mass than the galaxy would seem to contain. Dark matter was invented to resolve this discrepancy \cite{Zwicky_Masses_1937,Rubin_Rotational_1980}. For example, the data in Fig.~\ref{fig:rotation} can be explained by a spherical distribution of dark matter in the galactic halo \cite{van_Albada_1985}. Unfortunately, to date, the source of this extra mass has not yet been definitively identified\footnote{One interesting candidate, ``dressed neutrinos", is put forward in this volume by our friend and colleague, Yves Pomeau}. Also, there are countervailing view points. Milgrom \cite{Milgrom_MOND_1983} and others have argued that the observations should be interpreted as the need to modify Newton's third law at small accelerations; others, most notably Don Saari, have argued that the simple balance in terms of a net mass $M$ is not appropriate and that one must take account of forces between individual stars, and this may negate the need for adding any other matter.

Our starting point is the regularized Cross-Newell equation~\eqref{CNreduced}. Because the relaxation dynamics will drive the preferred wavenumber $k$ close to $k_0$, here unity, we can write the dynamics as a gradient flow for the effective (averaged) energy
\begin{equation}
\overline{E} = \int \left[\left(|\nabla\psi|^2 -1 \right)^2 + \epsilon^2 \left(\nabla^2\psi\right)^2\right]  dxdy, 
\label{CNred}
\end{equation}

We can write down a Lorentz-invariant generalization of the pattern Lagrangian~\eqref{CNred}  by analogy as 
\begin{align}
\mathcal{L}_P & = \rho_0 c^2   \int \left\{(|\nabla \psi|^2-c^{-2}\psi_t^2-1)^2 +  k_0^{-2} (\Delta \psi-c^{-2} \psi_{tt} )^2\right\} \ d^3x \,dt \label{Lagrangian} 
\end{align}
 where $\rho_0 c^2$ is the effective ``energy density" of the phase field and $k_0$ is its preferred wave-number is an inverse length. 
 
Although we do not have a concrete identification of the nature of the field $w$ nor the mechanism which leads to the local periodic structure $w = f(\psi), \nabla \psi = \mathbf{k}$, nor what sets the preferred wave-number $k_0$, we observe that~\eqref{Lagrangian} gives the {\em natural} Lorentz-invariant generalization of the {\em universal averaged energy} for nearly periodic stripe patterns~\eqref{CNred}, and is thus expected to describe the macroscopic behavior arising from a variety of microscopic models. The integral is set up with extra normalizing factors of $k_0$ so that $\bar{E}$ has the dimensions of action, i.e. energy integrated in time. 

The Euler-Lagrange equation for the Lagrangian in~\eqref{Lagrangian} is the 4th-order nonlinear wave equation
$$
(\Delta  - c^{-2} \partial_{tt})^2 \psi - 2 \eta^{\alpha \beta} \partial_\alpha\left\{(|\nabla \psi|^2-c^{-2}\psi_t^2-k_0^2) \partial_\beta \psi\right\} = 0
$$
on Minkowski spacetime with coordinate $(x^0= ct, x^1=x,x^2=y,x^3=z)$, (inverse) metric $\eta^{\alpha \beta}$ and signature $(- \,+\, +\, +)$. We seek stationary spherical target solutions $\psi = \psi(r)$ which both reflect the galactic halo and are ``localized", so that wavenumber mismatch $|\nabla \psi| - k_0$ vanishes as $r \to \infty$. 
For radial solutions $\psi = \psi(r) = \psi(\sqrt{x^2+y^2+z^2})$,  we have
\begin{equation}
\left(\partial_{rr} + \frac{2}{r} \partial_r\right)^2 \psi(r) - \frac{2}{r^2} \partial_r \left\{ r^2( \psi'(r)(\psi'(r)^2 - k_0^2) )\right\} = 0
\label{eleqn}
\end{equation}
It is immediately obvious that $\psi(r) = k_0 r$ is a solution, although this solution is not smooth at the origin. Nonetheless, we expect $\psi(r) = k_0r$ gives the correct far field behavior. Linearizing about this solution, $\psi(r) = k_0 r + \alpha (k_0 r)^\beta$, we get
$$
\frac{\alpha  k_0^\beta \beta  (\beta +1) \left(\beta ^2-3 \beta -4 k_0^2
   r^2+2\right)}{r^4}  = 0
   $$
   so that the far field corrections correspond to $\beta = 0$, a constant shift in $\psi$ and $\beta = -1$ corresponding to a correction of the form $\alpha/(k_0 r)$.  Eq.~\eqref{eleqn} is 4th order, so there are two other solutions,  one diverging exponentially as $r \to \infty$. Consequently we get one boundary conditions by requiring that $\psi'$ should stay bounded as $r \to \infty$.  At the origin regularity of the solution requires $\psi'(0) = \psi'''(0) = 0$. We are also free to pick $\psi(0)$ and this gives us a full complement of boundary conditions.  Expanding $\psi = \sum c_k (k_0 r)^{2k}$ in even powers of $k_0r$ and substituting in ~\eqref{eleqn} shows that $c_0$ is undetermined and the higher coefficients satisfy algebraic relations such that if $c_2$ is specified, then all the higher order coefficients are determined. The condition that $\psi'$ should remain bounded as $r \to \infty$ will now determine $c_2$.
   
   While we can, in principle, determine the full stationary target pattern solution, {\em e.g.} by numerically solving~\eqref{eleqn}, our interest is in the far field behavior of the stationary solution. From the preceding disscussion we have
   \begin{equation}
\psi \approx \begin{cases} k_0 r +  \psi_0 + \frac{\alpha}{k_0 r} + \frac{\alpha^2}{2 k_0^3 r^3} + O((k_0 r)^{-5})& r \to \infty \\ c_0 + c_2(k_0 r)^2 -\frac{c_2}{10} (k_0r)^4 + O((k_0 r)^6) & r \to 0 \end{cases}
\label{full-solution}
\end{equation} 
Again, we emphasize that, the only freedom left is an overall constant shift so that $\psi - c_0, \alpha$ and $c_2$ can be determined in principle, although we do not need their precise values for our purposes. 

For negative $\alpha$, the far-field wavenumber will be within the Busse balloon (the stable range of wavenumbers for the non-regularized Cross-Newell equation). 
Using the Ostrogradski formalism \cite{urries1998formalism} we can compute the Hamiltonian (or equivalently the mass) density \footnote{It follows from Ostrogradski's theorem that the Hamiltonian is {\em not bounded} from below. This is a serious concern, and our pattern states are necessarily unstable. We sidestep this issue by only considering stationary patterns in this work, but this issue needs to be revisited in a time dependent framework.}  corresponding to the Lagrangian~\eqref{Lagrangian} for the stationary solution~\eqref{full-solution}. In the far field, $R \gg k_0^{-1}$, the equivalent mass $M_P(R)$ in the pattern field $\psi$ is 
\begin{equation}
\label{dark-matter}
M_P(R)  = \frac{32 \pi \rho_0 (1-\alpha) R}{k_0^2} + \text{ small corrections}
\end{equation}
Matching the gravitational acceleration from this ``effective mass" in the pattern field, added to the visible {\em baryonic mass} $M_B(R)$,  with the centripetal acceleration $v^2/R$ of a circular orbit gives
\begin{equation}
\label{azimuthal}
\frac{v^2}{R} = \frac{G (M_B(R) + M_P(R))}{R^2} \quad \implies \quad v(R) \approx \left[\frac{G M_B(R)}{R} +  \frac{32 G \pi \rho_0(1-\alpha)}{k_0^2} \right]^{1/2},
\end{equation}
demonstrating the flattening of the rotation curve of a galaxy as $R$ gets large assuming $M_B$ is bounded.

We  will now sharpen this heuristic argument. The Stress-Energy-Momentum $T^{\alpha \beta}$ 
in the pattern field with Lagrangian $\mathcal{L}_P$ is given by the Einstein-Hilbert prescription $T^{\alpha \beta} = \frac{2}{\sqrt{-g}} \frac{\delta \tilde{L}}{\delta g_{\alpha \beta}}$  where $g = \mathrm{det}[g_{\alpha \beta}]$ and $\tilde{L}$ is the appropriate extension of the Lagrangian $\mathcal{L}_P$ in~\eqref{Lagrangian} to curved space-times \cite{MTW}.  The curved space Lagrangian can be obtained from the `minimal coupling' assumption \cite{MTW} as  
\begin{equation}
\tilde{L} = \frac{\rho_0 c^2}{k_0^4}   \int \left\{(\nabla^\mu \psi \nabla_\mu \psi-k_0^2)^2 +  (\nabla^\mu \nabla_\mu \psi )^2\right\} \sqrt{-g} \ d^4x,
\label{curved}
\end{equation}
where the metric $g_{\alpha \beta}$ has signature $(- \,+\, +\, +)$, $\nabla_\mu$ is the corresponding covariant derivative. 
To obtain the stress tensor in a (background) flat space time, it suffices to consider variations $g_{\alpha \beta} = \eta_{\alpha \beta} + t \rho_{\alpha \beta}$ and compute all the variations to first order in $t$. To this end, we record the following relations for the inverse metric $g^{\gamma \delta}$, the Christoffel symbols $\Gamma^\sigma_{\gamma \delta}$ and the quantities that appear in the curved space Lagrangian $\tilde{L}$:
\begin{align*}
g^{\gamma \delta} & = \eta^{\gamma \delta} - t \, \eta^{\gamma \alpha}\eta^{\delta \beta} \rho_{\alpha \beta} + O(t^2) \\
\Gamma^\sigma_{\gamma \delta} & = \frac{t}{2}\, \eta^{\sigma \xi} \left[ \frac{\partial \rho_{\gamma \xi}}{\partial x^\delta} +  \frac{\partial \rho_{\delta \xi}}{\partial x^\gamma} - \frac{\partial \rho_{\gamma \delta}}{\partial x^\xi}\right]+ O(t^2) \\
\sqrt{-g} & = 1 + \frac{t}{2} \eta^{\alpha \beta} \rho_{\alpha \beta} + O(t^2) \\
\nabla^\mu \psi \nabla_\mu \psi & =  \eta^{\gamma \delta} \partial_\gamma \psi \partial_\delta \psi  - t \, \eta^{\gamma \alpha}\eta^{\delta \beta} \rho_{\alpha \beta} \partial_\gamma \psi \partial_\delta \psi + O(t^2) \\
g^{\gamma \delta} \nabla_\gamma \nabla_\delta \psi & = \Box \, \psi - t \, \eta^{\gamma \alpha}\eta^{\delta \beta} \rho_{\alpha \beta}  \partial_{\gamma} \partial_{\delta} \psi - \frac{t}{2}\, \eta^{\gamma \delta} \eta^{\sigma \xi} \left[ \frac{\partial \rho_{\gamma \xi}}{\partial x^\delta} +  \frac{\partial \rho_{\delta \xi}}{\partial x^\gamma} - \frac{\partial \rho_{\gamma \delta}}{\partial x^\xi}\right] \partial_\sigma \psi + O(t^2)
\end{align*}
The Einstein-Hilbert stress tensor is now given by
$$
\left. \frac{d}{dt} \tilde{L} \right|_{t=0} = \frac{1}{2} \int T^{\alpha \beta} \rho_{\alpha \beta} \,d^4x + \text{ boundary terms}.
$$
A straightforward but somewhat lengthy calculation now yields
\begin{align}
T^{\alpha \beta}  =  \frac{\rho_0 c^2}{k_0^4} & \left\{   -4(\eta^{\mu \nu} \partial_\mu \psi \partial_\nu \psi-k_0^2) \eta^{\sigma \alpha}\eta^{\tau \beta} \partial_\sigma \psi \partial_\tau \psi  - 4\, \Box \psi\, \eta^{\sigma \alpha}\eta^{\tau \beta} \partial_{\sigma} \partial_{\tau} \psi  \right. \nonumber \\
& + 2 (\eta^{\alpha \tau} \eta^{ \beta \sigma} + \eta^{\beta \tau} \eta^{\alpha \sigma} - \eta^{\alpha \beta} \eta^{\sigma \tau} ) \partial_\tau (\Box \psi \, \partial_\sigma \psi) \nonumber \\
& + \left. \eta^{\alpha \beta} \left[(\eta^{\mu \nu} \partial_\mu \psi \partial_\nu \psi-k_0^2)^2 + (\Box \psi)^2\right] \right\}
\label{stress-tensor}
\end{align}
We can now express the energy density $T^{\alpha \beta}$ for the (stationary) phase field~\eqref{full-solution} with respect to a normalized basis $\{\mathbf{e}_t, \mathbf{e}_r, \mathbf{e}_\theta, \mathbf{e}_\phi\}$ induced by (spatial) spherical polar coordinates $(r,\theta,\phi)$.

We decompose the stress tensor into two pieces, $T_s$ coming from the ``stretching energy"  with density $(|\nabla \psi|^2 - c^{-2} (\partial_t \psi)^2 - k_0^2))^{2}$ and $T_b$ from the bending energy $(\Box \, \psi)^2$. We compute these quantities for a stationary, radial solution $\psi = \psi(r)$ to get
\begin{equation}
T_s \sim  \frac{\rho_0c^2}{k_0^4} \begin{pmatrix}  \tau_1 & 0 & 0 & 0 \\
 0 & \tau_2 &
   0 & 0 \\
 0 & 0 &-\tau_1 & 0 \\
 0 & 0 & 0 & -\tau_1 \end{pmatrix} \qquad
 T_b \sim  \frac{\rho_0c^2}{k_0^4} \begin{pmatrix}  \tau_3 & 0 & 0 & 0 \\
 0 & \tau_4 &
   0 & 0 \\
 0 & 0 &-\tau_3 & 0 \\
 0 & 0 & 0 & -\tau_3 \end{pmatrix}
 \label{stress}
\end{equation} 
where 
\begin{align}
\tau_1 & = (\psi'(r)^2-k_0^2)^2 \nonumber \\
\tau_2 & = (3 \psi'(r)^2 +k_0^2)(\psi'(r)^2- k_0^2) \nonumber \\
\tau_3 & = -\psi ''(r)^2-\frac{2 \psi '(r) \left(r \psi'''(r)+4 \psi ''(r)\right)}{r} \nonumber \\
\tau_4 & = \frac{8 \psi '(r)^2}{r^2}+\psi ''(r)^2-2 \psi'''(r) \psi '(r)
\label{geometric}
\end{align}

We remark on the expected structure of $T_s$ and $T_b$, {\em viz.} the off-diagonal stresses should be zero from time-reversal and spherical symmetries of the solution $\psi$, and further $T^{00} = -T^{\theta \theta} = - T^{\phi \phi}$ since the metric has signature  $(- \,+\, +\, +)$, and $\psi$ is independent of $t, \theta$ and $\phi$. Finally, the quantities $\tau_1, \tau_2, \tau_3$ and $\tau_4$ are constrained by the conservation of Energy-Momentum $\nabla_\mu T^{\mu \nu} = 0$. A calculation shows that, as expected,  these 4 conditions reduce to just one constraint on $\psi(r)$, namely that $\psi$ should satisfy the Euler-Lagrange equation~\eqref{eleqn}. 

The stress tensor associated with the field $\psi$ will act as source for the curvature of space-time. The far-field, $r \gtrsim k_0^{-1}$, stress tensor is obtained from~\eqref{full-solution}:
$$
T =  \frac{\rho_0c^2}{k_0^4} \begin{pmatrix} \frac{4 \alpha ^2-4 \alpha }{r^4} & 0 & 0 &
   0 \\
 0 &\frac{4 \alpha ^2-4 \alpha
   }{r^4}+\frac{8 k_0^2-8 \alpha  k_0^2}{r^2} & 0 & 0 \\
 0 & 0 & \frac{4 \alpha -4 \alpha ^2}{r^4} &
   0 \\
 0 & 0 & 0 & \frac{4 \alpha -4 \alpha
   ^2}{r^4}  \end{pmatrix}  + O((k_0r)^{-6})
   $$

We consider a spherical galaxy\footnote{For simplicity. A similar procedure can also be  used for other mass distributions, {\em e.g.} disk galaxies.} consisting of  (non-relativistic) Baryonic matter (pressure is approximately zero) with stress tensor $T^{00} = \rho_B(r), T^{rr} = T^{\theta \theta} = T^{\phi \phi} = P = 0$ along with the stationary phase field $\psi(r)$ from above. Assuming that the space-time is close to being flat, we can compute the galactic rotation curves in the Newtonian approximation \cite{MTW}. In this approximation, the metric  is of the form
\begin{equation}
g = -c^2(1 + 2  \Phi(r)) dt^2 + (1+2 \Lambda(r)) dr^2 + r^2(d \theta^2 + \sin^2\theta d \phi^2).
\label{eq:newtonian-limit}
\end{equation}
Computing the Einstein curvature tensor $G^{\alpha \beta}$ (to lowest order in $\Phi$ and $\Lambda$) and relating it to the components of the stress tensor $T^{\alpha \beta}$, we get the following 3 equations from the $tt$, $rr$ and $\theta \theta$ or $\phi \phi$ components of Einstein's equations  \cite{MTW} $G^{\alpha \beta} = 8 \pi G T^{\alpha \beta}/c^4$
\begin{align}
\frac{\partial_r (r \Lambda(r))}{r^2} & = \frac{4 \pi G}{c^2} \left(\rho_B(r) - \frac{ 4 \rho_0 \alpha(1-\alpha)}{k_0^4 r^4}\right) + O(r^{-6})\nonumber \\
\frac{r \Phi'(r) - \Lambda(r)}{r^2} & = \frac{32 \pi G}{c^2} \rho_0 \left[ \frac{(1-\alpha)}{k_0^2 r^2} - \frac{\alpha(1-\alpha)}{2 k_0^4 r^4}\right] + O(r^{-6})\nonumber \\
\frac{\partial_r( r \Phi'(r) - \Lambda(r))}{r} & =  \frac{32 \pi G}{c^2} \rho_0 \frac{\alpha(1-\alpha)}{k_0^4 r^4} + O(r^{-6})
\end{align}
The third equation is redundant as a consequence of the Bianchi identity $\nabla_\mu G^{\mu \nu} = 0$ and the conservation of Energy-Momentum $\nabla_\mu T^{\mu \nu} = 0$. We can now solve this system to get
\begin{align}
M_B(r) & = \int_0^r 4 \pi \rho_B(s) s^2 ds, \nonumber \\
\Lambda(r) & = \frac{G M_B(r)}{c^2 r} + \frac{16 \pi G \rho_0 \alpha(1-\alpha)}{c^2}\left[ \frac{1}{k_0^4 r^2} -\frac{1}{k_0^2} \right]+ O(r^{-4}), \nonumber \\
\Phi(r) & = -\frac{G M_B(r)}{c^2 r} + \frac{32 \pi G \rho_0(1-\alpha)}{c^2k_0^4} \log(k_0 r) + O(r^{-2}),
\end{align}
In integrating the equation for $\Lambda$, we have made the (immaterial) choice of imposing $\Lambda(r) = GM_B(r)c^{-2}r^{-1}$ at $r = k_0^{-1}$ to set the value of the constant of integrations.

We are now in a position to determine the rotation curves. The velocity of an particle in a circular orbit in the ``equatorial plane" $\theta = \pi/2$ is given by
$$
v^\alpha = \frac{dx^\alpha}{d \tau} = \frac{d}{d \tau}\left( t, r, \theta,\phi\right) = \left(1 - O\left(\frac{r^2 \omega^2}{c^2}\right), 0, 0, \omega(r)\right)
$$
where $\omega(r)$ is its 
angular velocity 
and $\tau$ is proper time for this particle. Since such a particle is in geodesic motion, we have $\frac{d}{d\tau} v^\alpha + \Gamma^\alpha_{\beta \gamma} v^\beta v^\gamma = 0$. Computing the Christoffel symbols for the metric in~\eqref{eq:newtonian-limit} and substituting the expression for $v^\alpha$ we get
\begin{equation}
v_\phi(r) = r \omega(r) =  \left[\frac{G M_B(r)}{r} + \frac{32 \pi G \rho_0(1-\alpha)}{k_0^4}\right]^{1/2}
\label{orbital}
\end{equation}
Note that we have recovered the profile $v = \sqrt{v_B^2+v_D^2}$ where $v_B$ and $v_D$ are the orbital velocities inferred from the visible baryonic matter and a hypothesized dark matter with an isothermal profile $\rho_D(r) \sim r^{-2}$ respectively \cite{galactic-dynamics}.  Also, this expression agrees with the heuristic formula~\eqref{azimuthal}.

To use this formula to compute $v_\phi$ we need one further ingredient, namely the quantity $\frac{\rho_0(1-\alpha)}{k_0^4}$. We can choose $\rho_0$ and $k_0$ to yield agreement with Verlinde's work on emergent gravity \cite{verlinde2017}, by requiring that 
\begin{enumerate}
\item At the cross-over length scale $k_0^{-1}$, the two terms in the square root match.
\item At the scale $r \sim k_0^{-1}$, the accelerations satisfy $k_0 v_B^2 \sim k_0 v_D^2 \sim a_0 \sim c H$ where $a_0$ is the natural acceleration scale in MoND \cite{Milgrom_MOND_1983} and $H$ is Hubble's constant \cite{verlinde2017}.
\end{enumerate} 
Using the parameters for the Milky way $M_B \sim 6 \times 10^{10} M_{\odot}$ and Hubble's constant $H \approx 67.6$ km/s/mpc, we obtain $k_0^{-1} \approx 3$ kpc and $v_D \approx 220$ km/s which are in the right ranges for the central bulge and the flat part of the rotation curve respectively. We also get the estimate $\rho_0(1-\alpha) \approx 1.2 \times 10^{-21}$ kg/m$^3$, which is to within a few orders of magnitude of the cosmological constant $\Lambda$, suggesting, perhaps, a connection between the mechanisms that generate $\Lambda$ and the mechanisms that generate ``dark matter" in our model. This is one of the key predictions in recent work on emergent gravity \cite{verlinde2017}.

We might therefore ask: Could dark matter simply be a manifestation of the elastic energy of a patterned structure which foliates the universe, and whose energy, reflecting the deviation of the local wavenumber from its preferred value, depends only on what we see as detectable (visible and black hole)  matter? 

We have no way of answering that question definitively. There is no doubt that, in the absence of any idea about the microscopic description of the early universe, these suggestions are purely conjectural and lack a means of verification. However, we have introduced a new paradigm and the new picture has the virtue that it is potentially consistent with observations \cite{Rubin_Extended_1978}. Also, it avoids some of the small-scale inconsistencies between observations and  the existing cosmological constant/cold dark matter ($\Lambda$CDM) paradigm \cite{Navarro_Core_2000,Weinberg_CDM_2015}, since there is a natural ``core" radius $ k_0^{-1}$ in our model

\setcounter{equation}{0}
\section{Inflation, entanglement and dark energy in a pattern universe} \label{apndx:quantum-patterns}

In this section we investigate phenomena that arise on macroscopic scales as the parameters in a microscopic ``quantum pattern equation"  are varied. The key idea is that, starting with a microscopic description, a master equation which captures behaviors at all scales, the physics at successively larger scales is revealed, stage by stage, by averaging over the dominant scales at the previous level.  Fields which vary on very long times and distances can behave very differently to fields which have very short scale variations. Among the consequences is that there need be no contradiction between super-luminal  spreading of the influences of short scale fluctuations, i.e. entanglements, while the communication of ``information", a coarse-grained, large scale notion is limited by the speed of light \cite{Hooft_Equivalence_1988}. 

We will begin with the microscopic equation, closely related to the one we used in the previous section.
\begin{align}
\label{microscopic} 
\rho w_{tt}+\mu w_t = & \nabla^2(D \nabla^4+P \nabla^2+a)w +b((\nabla w)^2+2w \nabla^2 w) \\
& -4c_1w^3+2c_2(w(\nabla w)^2+w^2\nabla^2w). \nonumber
\end{align}
We have applied an extra Laplacian operator to the RHS of~\eqref{microscopic} as we want to include the effects of a Goldstone or soft mode which will turn out to parallel the Einstein cosmological ``constant" or quintessence. The stress parameter $P$ here will essentially parallel the inverse temperature in cosmological models. Alternatively, we can take $P$ fixed and let the stress parameter be $a$, which we will do here.  Nonlinear effects are included in the quadratic and cubic terms. For simplicity, we choose the RHS of~\eqref{microscopic} to be the variational gradient of a functional $E$, the energy,
\begin{equation}
\label{energy-univ}
E = \frac{1}{2}\int\left\{ D (\nabla(\nabla^2 w))^2 - P(\nabla^2 w)^2 + (a+bw)(\nabla w)^2 + 
c_1 w^4 + c_2 w^2 (\nabla w)^2\right\} d\mathbf{x},
\end{equation}
although, for most pattern behaviors, that restriction is not really necessary. Let us also think of the friction $\mu$ as being very small and, at least to start with, we take it to be zero.

For large $a$, equivalently small $P$, and for small amplitude values of the field $w$, the motion is wavelike, $w \sim \exp(i\mathbf{k\cdot x}-i \omega t)$,with dispersion relation, 
\begin{equation}
\label{dispersion}
\omega^2=k^2\left(D\left(k^2-\frac{P}{2D}\right)^2+a-\frac{P^2}{4D}\right).
\end{equation}

Thus, at the beginning, the $w=0$ state is neutrally stable and small perturbations behave as a sea of waves. A field of weakly nonlinear, dispersive waves will interact, share energy via resonances, and that energy will more and more be spread to the smallest scales. This is basically a fundamental result of wave turbulence theory (see references \cite{zakharov2012kolmogorov,newell2011wave}) and is consistent with the idea that the system can more effectively explore its phase space (increase entropy) by transferring energy to smaller and smaller scales. It is also known from wave turbulence theory that the smallest scales exhibit isotropy and homogeneity. The speed at which the spectral energy reaches the smallest scales depends of the competition between the relative strengths of the nonlinear terms and linear terms at large $k$. For situations with finite capacity such as fully developed hydrodynamic turbulence, the smallest scales are reached in finite time. For what are called infinite capacity situations, the time can be exponential. At small wavelengths, the energy spectrum is also isotropic. Because wave-packet speeds increase with increasing $k$, the size of an initial bubble of waves will rapidly expand either at an exponential or supra-exponential rate. An initial bubble containing a turbulent field of random waves spreads at inflationary rates and the small scale structures which dominate the outermost parts of this universe are isotropic.

As the size gets larger and larger, and the stress parameter $P$ grows, there comes a stage when an instability or phase transition occurs and certain shapes and scales are preferentially amplified. If we examine the linear stability of the $w=0$ solution by setting $w \sim \exp(i\mathbf{k \cdot x}+\sigma t)$, we find, analogous to~\eqref{dispersion}, that
\begin{equation}
\label{growthrate2} 
\sigma^2=-k^2\left(D\left(k^2-\frac{P}{2D}\right)^2+a-\frac{P^2}{4D}\right).
\end{equation}
The growth rate first becomes zero when 
\begin{equation}
\label{critcal} 
 a = a_c = \frac{P^2}{4D}, \quad k^2 =  k_c^2=\left(\frac{a}{D}\right)^{1/2}. 
\end{equation}

As $a$ decreases so that $a-P^2/4D$ becomes negative, the maximum amplification occurs for $a<a_c$ at scales $k_0=k_c = \left(\frac{a_c}{D}\right)^{1/4}$. Out of the random sea of, by now very weak nonlinear waves, a dominant periodic or quasi-periodic structure emerges. Rotational symmetry means of course that there are many competitors, namely all modes with wave-vectors with wavenumber $k_c$ and indeed their immediate neighbors. These modes compete for dominance via the nonlinear interactions and, eventually, a winner, or a set of possible winners depending on available symmetries, emerges. Often that winner can be a single mode, a pattern of stripes or rolls, with a wave-vector whose direction is arbitrary (and picked by local biases; i.e., symmetry breaking) but it can also be a combination of modes all with the same preferred wavenumber. Here, the mode is a stationary wave. Its group velocity, however, is finite.  

Moreover, for conservative systems such as~\eqref{microscopic} in the absence of friction, there are potentially other players which are not amplified but which are not damped either. The most important of these is the zero or constant mode. It represents a symmetry of the system in which an arbitrary real number is added to $w$  (analogous to the situation in a convecting fluid at small Prandtl numbers where the symmetry in the pressure field-the pressure only appears as a gradient-can drive large scale mean drift flows) is a Goldstone or soft mode and will play a very important role in the system's dynamics and in its choice of a final state.  In principle, we cannot neglect all the oscillatory modes either but we will argue here that once the first phase transition is reached, a friction which is zero at $k=0$ comes into play, and these other modes are damped. We will therefore seek to understand the evolution of the envelope $A$ of the most excited wave under self interactions and interactions with the Goldstone mode $B$. 

We insert the shape
\begin{equation}
\label{testshape} 
w=A e^{i \mathbf{k_0 \cdot x}}+\mathrm{cc}+B +\mathrm{corrections}, \quad \mathbf{k_0}^2 =\sqrt{\frac{a_c}{D}},
\end{equation}
into~\eqref{microscopic} and allow, in order to account for the fact that wave-vectors nearby to the preferred mode also play a role, both the envelope $A$ and the mean $B$ be slowly varying functions of space and time. Standard analysis \cite{newell1974envelope} leads to two equations for the evolution of both the envelope $A$ and the mean $B$. They take the form of two nonlinear wave equations,
\begin{align}
\rho \frac{\partial^2A}{\partial t^2}-4a_c (\mathbf{\hat{k}_0}\cdot \nabla)^2 A & = -2 b k_0^2AB + \text{growth, saturation and higher derivatives on $A,B$.} \label{eqA} \\
\rho \frac{\partial^2 B}{\partial t^2}-a_c \nabla^2 B& = -2bk_0^2AA^*+ \text{growth, saturation and higher derivatives on $A,B$.} \label{eqB}
\end{align}
Here $\mathbf{\hat{k}_0}$ is the unit vector in the direction of $\mathbf{k}_0$. 
The reason one does not get the familiar and canonical complex Ginzburg-Landau equation \cite{newell1974envelope} is that the frequency $\omega$ (alternatively  the growth rate $\sigma$) has a double zero at the onset value $k_0^2=\sqrt{\frac{a_c}{D}}$. In fact, properly taking the limit, the two wave speeds, $2\sqrt{a_c}$ and $\sqrt{a_c}$, are the respective (critical) group velocities of the modes.

The equations~\eqref{eqA}~and~\eqref{eqB} tell us how all information depending on the envelope of the amplified mode $A$ and the dynamically important but neutral mode $B$ propagate. We think of $c=2\sqrt{a_c}$ as the ``speed of light". 

We make two important points.
First, while information associated with functionals of the original microscopic field $w$ can potentially travel at any speed (there is therefore nothing to contradict the phenomenon of ``entanglement"), information which is a functional only of the envelope $A$ which varies on long scales travels with a definite speed $c$ determined by the microscopic parameters, in this case is $\sqrt{2}a$. 

Second, we regard the mean field $B$ as being driven by the dc component proportional to $AA^*$ arising from the quadratic term. It is not, in any point by point manner, proportional to $AA^*$ but rather captures the integrated history of the envelope intensity. The equation~\eqref{eqA} then parallels the Einstein equation (in the near flat universe approximation) which describes how the curvature of space-time is deformed by the presence of distributed mass. The term $AB$, again arising from the quadratic interaction in~\eqref{microscopic}, plays exactly the role that the cosmological constant term would play in orthodox theory. Whereas the function $B$ is not constant, it reflects some average over the state of the intensity of the amplified fluctuations. In a pattern universe, therefore, a Goldstone mode plays the role of a cosmological constant accelerating the expansion of the universe.

Although it may not be general, there is an interesting special case of equations \eqref{eqA}, \eqref{eqB}. The spatial gradient term in \eqref{eqA} is the scalar product of the vector group velocity of the most unstable wave and the gradient. It is well known that, in three wave systems, the condition of three wave resonance between long (here $B$) and short waves (here $A$) reduces to the statement that when the projection of the group velocity of the latter is equal to the phase velocity (here also its group velocity) of the former. This occurs when $\mathbf{\hat{k}_0}$ is $(1/2,\pm \sqrt{3}/2)$. In such circumstances the response of the ``long'' wave (here $B$) to the short wave envelope (here $A$) is very strong so that as the intensity of the short wave decreases, the strength of the ``long'' wave increases. If the ``long'' wave were to represent the cosmological constant, then it would get stronger as the gravity induced deviation from the Minkowski metric decreased, thus accelerating the expansion.

It is also worth thinking about the analogy with the dynamics of the complex Ginzburg-Landau equation and its purely dispersive first cousin, the nonlinear Schrodinger (NLS) equation, both similar to the order parameter equations of this model. In the case of the NLS, the equivalent hydrodynamic description obtained by setting $\psi$, the variable satisfying NLS, equal to $\sqrt{\rho} \exp(i \phi)$. Up to an extra term called the quantum pressure, the two corresponding equations for $\rho$ and $\vec{v}= \nabla \phi$ are the Euler equations for the flow of a fluid with density $\rho$ and a pressure proportional to either $\mp \rho$ depending on whether the NLS (there is a corresponding condition for the CGL equation) is focusing or defocusing. In the focusing case, the pressure increases as $\rho$ decreases, the key signature of the role of dark energy.

In summary, then, our picture is that, before the first phase transition, the field consists of a bubble of rapidly expanding nonlinear waves. At the first phase transition, the microscopic field $w$ develops a periodic structure. Even though all information about the behavior of the envelope $A$ is contained in~\eqref{microscopic}, its behavior is dominated by the pattern averaged equations and the physics associated with long scales. Subsequent phase transitions may introduce further separations of behaviors, each associated with an envelope of the field of the most excited mode at that transition. There is no reason that the scales of the later transitions should not lie between the ``Planck" scale and the very long scales on which the equation for the envelope $A$ obtains. 

Now what happens as $P$, the stress, continues to increase (the temperature continues to decrease)? We know from previous analyses on pattern formation \cite{Passot_Towards_1994} that, while near onset the point defects have to have surrounding vector rather than director (spinor) fields, far from onset the amplitude of $A$ is determined by the gradients of its phase and that phase can be double valued. In other words, near onset the order parameter $A$ is complex. Both the amplitude and its phase (its phase gradient is the vector $k$) are active order parameters, each obeying a dynamics governed by the solvability equation~\eqref{eqA}. For ease of visualization, let us consider the case of two space dimensions. In that case, the only point defects are composites of the concave and convex disclinations discussed in appendix~\ref{apndx:patterns}. They are saddles, vortices, targets, dislocations. However, as $P$ increases away from its critical onset value, the amplitude of the envelope becomes algebraically slaved to its phase gradient (and, because of rotational symmetry, on the modulus $k$ of the wave-vector $\mathbf{k}$). There is then only one real order parameter and the direction of the wave-vector $\mathbf{k}$ cannot be determined from the original microscopic field $w$ and its gradient. This point was previously emphasized in appendix~\ref{apndx:patterns}. It is at this stage that pattern quarks and pattern leptons can appear. They are unbound. Indeed, composites are generically unstable as was shown clearly in \cite{Passot_Towards_1994}.  At the next phase transition, however, when the relevant order parameter again becomes complex, and therefore the phase gradient $\mathbf{k}$ is a vector, pattern quarks and leptons begin to merge to form bound states. In  \cite{Passot_Towards_1994}  we showed that, when we begin with a field containing isolated concave and convex disclinations, and then lower the stress $P$ back towards it critical value $P_c$, the isolated disclinations disappear to be replaced by composites. This next transition might then be considered as the analogue of the separation of the grand unified theory into the strong and electroweak forces. An even later transition would parallel the separation of the weak and electromagnetic forces.
 
 In summary, we have made a prima facie case that pattern or crystal universes can exhibit many features which have analogues to that vast array of behaviors observed in fundamental particle physics and cosmology, most of which, dark matter, dark energy, are as yet unexplained by orthodox theories. Our basic idea has been that the pattern universe is one containing structures at many scales, each set being determined by the preferred configurations arising at a set of discrete phase transitions. We have shown that objects with some of the same invariants as quarks and leptons arise naturally without any imposition of symmetries that might induce particular fractional values. We have shown that a universe foliated by phase surfaces can give rise to additional gravitational like forces which lead to star rotation speeds consistent with observations. We have explained how there can be a rapid expansion of the early pattern universe leading to an isotropic far field. We have suggested that the nature of the field which promotes an accelerated universe is related to the presence of Goldstone mode released at a phase transitions. In short, we have offered a new paradigm the ultimate value of which will only be known after  further work.

\section*{Acknowledgments}
ACN was supported  by the NSF through the award DMS 1308862 and SCV was supported by the Simons Foundation through awards 524875 and 560103. We are grateful to Yves Pomeau and Amit Acharya for many stimulating discussions.  Indeed, in this volume, Yves Pomeau has also suggested an intriguing candidate for dark matter in the form of a ``dressed" neutrino. Portions of this work were carried out when SCV was visiting the Center for Nonlinear Analysis at Carnegie Mellon University.


\begin{thebibliography}{10}

\bibitem{boomerang}
P.~de~Bernardis, P.~A.~R. Ade, J.~J. Bock, J.~R. Bond, J.~Borrill,
  A.~Boscaleri, K.~Coble, B.~P. Crill, G.~De~Gasperis, P.~C. Farese, P.~G.
  Ferreira, K.~Ganga, M.~Giacometti, E.~Hivon, V.~V. Hristov, A.~Iacoangeli,
  A.~H. Jaffe, A.~E. Lange, L.~Martinis, S.~Masi, P.~V. Mason, P.~D. Mauskopf,
  A.~Melchiorri, L.~Miglio, T.~Montroy, C.~B. Netterfield, E.~Pascale,
  F.~Piacentini, D.~Pogosyan, S.~Prunet, S.~Rao, G.~Romeo, J.~E. Ruhl,
  F.~Scaramuzzi, D.~Sforna, and N.~Vittorio.
\newblock A flat {U}niverse from high-resolution maps of the cosmic microwave
  background radiation.
\newblock {\em Nature}, 404:955 EP --, 04 2000.

\bibitem{wmap}
G.~Hinshaw, D.~Larson, E.~Komatsu, D.~N. Spergel, C.~L. Bennett, J.~Dunkley,
  M.~R. Nolta, M.~Halpern, R.~S. Hill, N.~Odegard, L.~Page, K.~M. Smith, J.~L.
  Weiland, B.~Gold, N.~Jarosik, A.~Kogut, M.~Limon, S.~S. Meyer, G.~S. Tucker,
  E.~Wollack, and E.~L. Wright.
\newblock Nine-year {W}ilkinson {M}icrowave {A}nisotropy {P}robe ({WMAP})
  {O}bservations: {C}osmological {P}arameter {R}esults.
\newblock {\em The Astrophysical Journal Supplement Series}, 208(2):19, 2013.

\bibitem{Peebles_The_2003}
P.~J.~E. Peebles and B.~Ratra.
\newblock The cosmological constant and dark energy.
\newblock {\em Rev. Mod. Phys.}, 75:559--606, Apr 2003.

\bibitem{Rubin_Rotation_1970}
V.~C. {Rubin} and W.~K. {Ford}, Jr.
\newblock {Rotation of the Andromeda Nebula from a Spectroscopic Survey of
  Emission Regions}.
\newblock {\em Astrophysical Journal}, 159:379, February 1970.

\bibitem{newell2012pattern}
Alan~C. Newell.
\newblock Pattern quarks and leptons.
\newblock {\em Applicable Analysis}, 91(2):213--223, 2012.

\bibitem{newell2014quarks}
Alan~C. Newell.
\newblock `quarks' and `leptons' in three dimensional patterns.
\newblock {\em European Journal of Mechanics - B/Fluids}, 47:39 -- 47, 2014.
\newblock Enok Palm Memorial Volume.

\bibitem{NV18}
Alan~C. Newell and Shankar~C. Venkataramani.
\newblock Elastic sheets, phase surfaces, and pattern universes.
\newblock {\em Studies in Applied Mathematics}, 139(2):322--368, 2018.

\bibitem{newell1996defects}
A.~C. Newell, T.~Passot, C.~Bowman, N.~Ercolani, and R.~Indik.
\newblock {Defects are weak and self-dual solutions of the Cross-Newell phase
  diffusion equation for natural patterns}.
\newblock {\em Physica D: Nonlinear Phenomena}, 97(1):185--205, 1996.

\bibitem{Blumenthal_Formation_1984}
George~R. Blumenthal, S.~M. Faber, Joel~R. Primack, and Martin~J. Rees.
\newblock Formation of galaxies and large-scale structure with cold dark
  matter.
\newblock {\em Nature}, 311(5986):517--525, 10 1984.

\bibitem{Davis_Evolution_1985}
M.~{Davis}, G.~{Efstathiou}, C.~S. {Frenk}, and S.~D.~M. {White}.
\newblock {The evolution of large-scale structure in a universe dominated by
  cold dark matter}.
\newblock {\em Astrophysical Journal}, 292:371--394, May 1985.

\bibitem{Zwicky_Masses_1937}
F.~{Zwicky}.
\newblock {On the Masses of Nebulae and of Clusters of Nebulae}.
\newblock {\em Astrophysical Journal}, 86:217, October 1937.

\bibitem{Rubin_Rotational_1980}
V.~C. {Rubin}, W.~K. {Ford}, Jr., and N.~{Thonnard}.
\newblock {Rotational properties of 21 SC galaxies with a large range of
  luminosities and radii, from NGC 4605 (R = 4kpc) to UGC 2885 (R = 122 kpc)}.
\newblock {\em Astrophysical Journal}, 238:471--487, June 1980.

\bibitem{Rubin_Extended_1978}
V.~C. {Rubin}, N.~{Thonnard}, and W.~K. {Ford}, Jr.
\newblock {Extended rotation curves of high-luminosity spiral galaxies. IV -
  Systematic dynamical properties, SA through SC}.
\newblock {\em Astrophysical Journal Letters}, 225:L107--L111, November 1978.

\bibitem{van_Albada_1985}
T.~S. {van Albada}, J.~N. {Bahcall}, K.~{Begeman}, and R.~{Sancisi}.
\newblock {Distribution of dark matter in the spiral galaxy NGC 3198}.
\newblock {\em Astrophysical Journal}, 295:305--313, August 1985.

\bibitem{Milgrom_MOND_1983}
M.~{Milgrom}.
\newblock {A modification of the Newtonian dynamics as a possible alternative
  to the hidden mass hypothesis}.
\newblock {\em Astrophysical Journal}, 270:365--370, July 1983.

\bibitem{urries1998formalism}
F.~J. de~Urries and J.~Julve.
\newblock Ostrogradski formalism for higher-derivative scalar field theories.
\newblock {\em Journal of Physics A: Mathematical and General}, 31(33):6949,
  1998.

\bibitem{MTW}
C.~W. {Misner}, K.~S. {Thorne}, and J.~A. {Wheeler}.
\newblock {\em {Gravitation}}.
\newblock W.H.~Freeman and Co., San Francisco, 1973.

\bibitem{galactic-dynamics}
James Binney and Scott Tremaine.
\newblock {\em Galactic dynamics}.
\newblock Princeton Univ. Press, 2008.

\bibitem{verlinde2017}
Erik~P. Verlinde.
\newblock {Emergent Gravity and the Dark Universe}.
\newblock {\em SciPost Phys.}, 2:016, 2017.

\bibitem{Navarro_Core_2000}
Julio~F. Navarro and Matthias Steinmetz.
\newblock The core density of dark matter halos: a critical challenge to the
  {$\Lambda$CDM} paradigm?
\newblock {\em The Astrophysical Journal}, 528(2):607, 2000.

\bibitem{Weinberg_CDM_2015}
David~H. Weinberg, James~S. Bullock, Fabio Governato, Rachel Kuzio~de Naray,
  and Annika H.~G. Peter.
\newblock Cold dark matter: Controversies on small scales.
\newblock {\em Proceedings of the National Academy of Sciences},
  112(40):12249--12255, 10 2015.

\bibitem{Hooft_Equivalence_1988}
Gerard '\MakeLowercase{t}~Hooft.
\newblock Equivalence relations between deterministic and quantum mechanical
  systems.
\newblock {\em Journal of Statistical Physics}, 53(1):323--344, 1988.

\bibitem{zakharov2012kolmogorov}
Vladimir~E Zakharov, Victor~S L'vov, and Gregory Falkovich.
\newblock {\em Kolmogorov spectra of turbulence I: Wave turbulence}.
\newblock Springer Science \& Business Media, 2012.

\bibitem{newell2011wave}
Alan~C. Newell and Benno Rumpf.
\newblock Wave turbulence.
\newblock {\em Annual review of fluid mechanics}, 43:59--78, 2011.

\bibitem{newell1974envelope}
Alan~C. Newell.
\newblock Envelope equations.
\newblock In {\em Nonlinear wave motion ({P}roc. {AMS}-{SIAM} {S}ummer {S}em.,
  {C}larkson {C}oll. {T}ech., {P}otsdam, {N}.{Y}., 1972)}, pages 157--163.
  Lectures in Appl. Math., Vol. 15. Amer. Math. Soc., Providence, R.I., 1974.

\bibitem{Passot_Towards_1994}
T.~Passot and A.~C. Newell.
\newblock Towards a universal theory for natural patterns.
\newblock {\em Physica D: Nonlinear Phenomena}, 74(3--4):301 -- 352, 1994.

\end{thebibliography}

\end{document}